\documentclass{article}
\usepackage[utf8]{inputenc}
\usepackage{graphicx}
\usepackage{times}
\usepackage{xcolor}
\usepackage{amsmath}
\usepackage{ifpdf}
\usepackage{subfigure}
\usepackage{amssymb}
\usepackage{epstopdf}
\usepackage[numbers,sort&compress]{natbib}
\usepackage{appendix}
\usepackage{multirow}
\newcommand{\be}{\begin{equation}
\newcommand{\ee}{\end{equation}}}
\newcommand{\bea}{\begin{eqnarray}}
\newcommand{\eea}{\end{eqnarray}}
\newcommand{\nn}{\nonumber}
 \newcommand{\ev}[1]{\left\langle #1 \right\rangle}
\begin{document}

\title{Hard confinement of a two-particle quantum system using the variational method} 

\author{Nataly Rafat sabbah$^{1}$, Mohamed Ghaleb Al-Masaeed$^{2,3}$, Ahmed Al-Jamel$^{1}$   \\
$^1$Physics Department, Faculty of Science, Al al-Bayt University,\\ P.O. Box 130040, Mafraq 25113, Jordan\\
$^2$Ministry of Education, Jordan\\ $^3$Ministry of Education and Higher Education, Qatar\\
sabbahnataly@gmail.com\\moh.almssaeed@gmail.com \\aaljamel@aabu.edu.jo, aaljamel@gmail.com\\\\}

\maketitle

\begin{abstract}
The variational method is used to study hard confinement of a two-particle quantum system in two potential models, the Cornell potential and the global potential, with Dirichlet type boundary condition at various cut-off radii. The trial wavefunction is constructed as the product of the $1S$ free hydrogen atom wavefunction or $1S$ free harmonic oscillator wavefunction times a cut-off function of the form $(r-z)$ to ensure hard entrapment within a sphere of radius $z$. The behavior of $|\psi|^2$, the wavefunction at the origin (WFO), and the mean radius $\ev{r}$ are computed for different situations and compared for the two potential models.
\\

\textit{Keywords:}  Confined systems, variational method, Cornell potential, global potential. 
\end{abstract}

\section{Introduction}
Confined systems are very useful models in various fields of physics, ranging from nanophysics to high energy physics, for describing various effects. Historically, a confined hydrogen atom in a spherical box was firstly investigated by Michel’s et al in 1937 \cite{michels1937remarks} to study the influence of extremely high pressure on the atomic and molecular polarizabilities. Another perspective of confined atoms is due to their presence inside a solid. Confinement of a system is usually described in terms of a phenomenological potential model \cite{bugg2012hadron}. Within the framework of nonrelativistic quantum mechanics, the energy spectra as well as other physical properties due to confinement can be found by solving the Schrodinger equation subject to Dirichlet type bounding condition. There are various methods for doing this such as numerical methods, the Laplace transform, and other approaches. Sometimes, the analytic solution of the Schr\"odinger equation is not possible, and rather we use approximation methods. One of these, is the variational method, which is used to compute the ground state energies for the system under study using several suitable potentials, and then we compare our results from different methods. There are numerous benefits of has a simple form if there is just one parameter in the trial wavefunction. Then, in terms of actual application and physical discussion, it is really useful \cite{ding1999variational}. Many studies have been produced on the subject, using a variety of potential models. The researchers chose certain phenomenological potentials to examine confined systems' static features, such as their mass spectra; for example, Cornell, Martin, and logarithmic potentials \cite{feili2011calculation}. Furthermore, One-body integrals about Gaussian confinement potentials over Gaussian basis functions have been calculated and implemented \cite{https://doi.org/10.1002/qua.27043}. 
 \\

The atoms H, He, C, and K as well as the molecule $CH_4$ at the center of a spherical soft confinement potential of the type $V_N(r)=(\frac{r}{r_0})^N$ with stiffness parameter N and confinement radius $r_0$ have all been thoroughly studied non-relativistically. The soft confinement potential provides a more comprehensive view of spherical confinement as it approaches the hard-wall limit as $N \to \infty$ \cite{pavsteka2020atoms}. Besides, The characteristics of atomic systems with one electron that are spherically symmetric have been examined, specifically for those in which the electron is positioned between two impermeable concentric spheres with radii of $R_{int}$ and $R_{ext}$ \cite{https://doi.org/10.1002/qua.25887}.
 In Ref \cite{https://doi.org/10.1002/qua.20220} have demonstrated that the exact modified form of the virial theorem for restricted systems can be obtained using two straightforward semiclassical strategies: one based on the Wilson-Sommerfeld rule and the other on the uncertainty principle. A simplified, alternate quantum mechanical path to this outcome is also outlined. Key trends are revealed by pilot calculations on restricted oscillators.
\\

In 2015, An approximate recipe for calculating the energy eigenvalues of a non-relativistic hydrogen-like atom confined in a spherical cavity with an impenetrable wall was presented, using an Ansatz solution with a cut-off function and applying the Nikiforov-Uvarov method to solve an algebraic equation for the energy \cite{al2015energy}. Moreover, in Ref \cite{al2019heavy}, the study used the Nikiforov-Uvarov method to solve the radial Schrödinger equation for heavy quarkonia confined within a potential that accounts for conformal symmetry and finds that a small perturbing term in the potential achieves good agreement with experimental data.  The trial wavefunctions are constructed as the product of a wavefunction, similar to the wavefunctions of these for the ground state hydrogen or harmonic oscillator, times a cut-off function to ensure the vanishing of the whole wavefunction at the boundaries. This type of constraint could be called hard confining in contrast to soft confining where the system is confined but with a wavefunction that vanishes at infinity rather than at a certain finite radius.

In this paper, we study some properties of hard confinement of two-particle quantum systems using two potential models of great interest in physics, ranging from molecular, atomic, and particle physics. For instance, the first is the Cornell potential, which is one of the most common potential models in studying heavy meson mass spectra \cite{eichten1975spectrum,eichten1976interplay,godfrey1985mesons,hall2015schrodinger,soni2018q,chen2013spectral}.  It has a Coulomb-plus-linear form and has been used to investigate heavy quarkonium systems, serving as the first potential model for such studies and inspiring other phenomenological models \cite{mutuk2019cornell}.  The second potential model is known as global potential by some researchers \cite{brown2004quantum,motyka1995optimization,zalewski1998nonrelativistic}, as an alternative model to the Cornell potential. The non-Coulombic part of these potentials can be used to model confinement in other quantum systems as well. The variational method will be used, with trial wavefunctions containing a linear cut-off function, to find various physical properties of the system and the effect of the radius of confinement on these properties.
\section{Theoretical framework}
The main purpose of this study is to investigate the properties of hard confinement of a two-particle quantum system using variational method. Here, hard confinement refers to the situation where the system is confined within an impenetrable spherical box of finite radius $z$ with the possibility of interaction by a potential only inside that box. Here, we consider the following potential models of interest in physics, namely, the Cornell potential and the global potential.
\subsection{Potential models}
The Cornell potential, that is a Coulomb-plus-linear potential, has received a great deal of attention as an important non-relativistic model for the study of heavy quarkonium systems and inspired other phenomenological models.  The Cornell potential reads as \cite{mutuk2019cornell}:
\be
\label{cor}
V(r)=- \frac{A}{r}+ B r, \quad      r > 0,
\ee
where the potential parameters $A$ and $B$ have different numerical values, depending on the problem under investigation.  The first term is a Coulomb-like term, and the second term is a linear term that reflects confinement. Within the context of the heavy quarkonium problem, the Coulomb-like term arises from one-gluon exchange, and the linear term presumably arises from higher-order effects \cite{mutuk2019cornell}.

\begin{figure}[htb!]
    \centering
    \includegraphics[scale=0.8]{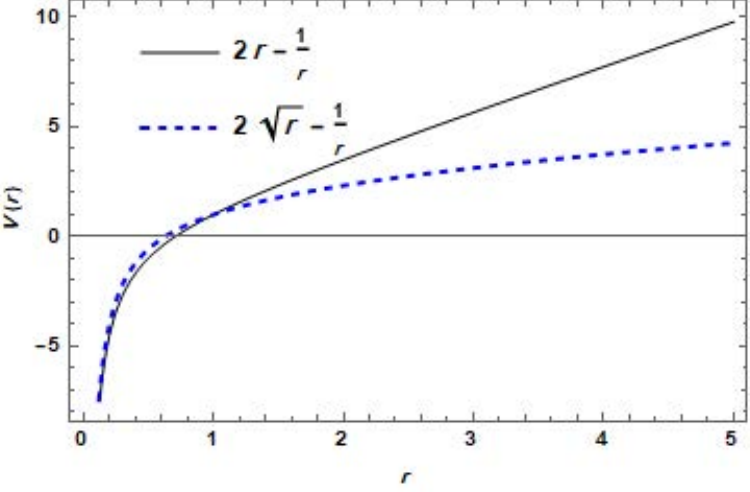}
    \caption{The two potential models used in this work for certain potential parameters.}
    \label{fig:my_label}
\end{figure}

The global potential takes the form \cite{motyka1995optimization}
\be
V(r)=(A\sqrt{r}-\frac{B}{r})+C,      
\ee
where $A,B,C$ are the potential parameters. This potential is used by some researchers to model the interaction between quarks and anti-quarks. Figure~(\ref{fig:my_label}) shows the two potential models for some potential parameters.\newpage
 Both potentials have the same behavior at very small distances, asymptotic freedom, as well as at large distances where confinement is dominant.
\subsection{Trial wavefunctions}
For problems with the previous potential models, a general form for the trial wavefunction for  1st state takes  the form
\be
\psi_{trial}(r)= N e^{-a r^b}.                                        
\ee
Here, $N$ is a normalization constant, $a$ is the variational parameter and $b$ is a constant related to the problem. The form of the trial wavefunction is determined by the model parameter ‘b’. In practice, four trial wavefunctions for the corresponding problems, but with soft confinement, were proposed \cite{feili2011calculation}:
\begin{enumerate}
    \item $b = 1$, namely $\psi_{trial}(r)= N e^{-a r}$. This corresponds to the $1S$ hydrogen wavefunction or exponential wavefunction.
    \item $b = 2$, namely $\psi_{trial}(r)= N e^{-a r^2}$. The harmonic oscillator wavefunction or a Gaussian wavefunction.
    \item $b = \frac{3}{2}$, namely $\psi_{trial}(r)= N e^{-a r^{\frac{3}{2}}}$. 
    \item $b = \frac{4}{3}$, namely $\psi_{trial}(r)= N e^{-a r^{\frac{4}{3}}}$.
\end{enumerate}
In this study, we take the first two model parameters $b = 1$ and $b = 2$, and we select a wavefunction that is appropriate for hardly confined systems in order to determine the ground state properties. We take then the ground-state trial wavefunction of the form:
\be
\psi_{trial}(r)= N (r-z) e^{-a r^b},                                      
\ee
where $z$ is the cut-off radius and $N$ is the normalization constant, which can be calculated from the normalization condition:
\be
\int |\psi_{trial}(r)|^2 d^3r = 1.
\ee
Then,
\be
\label{6}
\frac{1}{N^2} = 4 \pi \int_0^z r^2 (r-z)^2  e^{-2a r^b} dr. 
\ee
The wavefunction at the origin (WFO) is
\be
|\psi(0)|^2 = N^2 z^2.
\ee
The expectation value of the Hamiltonian of the system is
\be
\ev{H}=\ev{T}  + \ev{V},
\ee
where $\ev{V}$ is the expectation value of potential energy given as:
\be
\ev{V} = 4 \pi \int_0^z   \psi^*(r) \psi(r) V(r) r^2 dr, 
\ee
and $\ev{T}$ is the expectation value of the kinetic energy given as:
\be
\ev{T} = 4 \pi \int_0^\infty  \psi^*(r) * r^2 \hat{T} \psi(r)  dr,
\ee
where $\hat{T}$ is the kinetic energy operator:
\be
\hat{T}= - \frac{1}{2\mu} \left( \frac{2}{r} \frac{\partial}{\partial r} +  \frac{2}{r} \frac{\partial^2}{\partial r^2}\right),
\ee
where $\mu$ is the reduced mass of the system. To calculate $\ev{H}$, we must specify the wavefunction. We choose two trial forms\\
1. Hydrogen wavefunction or exponential wavefunction $(b = 1)$
 \be
\psi(r)= N(r-z) e^{-a r},  
 \ee
 2. Harmonic oscillator wavefunction $(b = 2)$
 \be
\psi(r)= N(r-z) e^{-a r^2},  
 \ee
where  $a>0,  z>0$. Once all needed quantities are determined, the variational method from which we obtain the values of a that minimizes $H$  for certain input potential parameters 
\be
\frac{\partial H}{\partial a}=0
\ee
\section{Derivations and Calculations}
In the following, we use the natural units with $\mu=1$GeV.
\subsection{For the Cornell potential}
For the Hydrogen wavefunction or exponential wavefunction $(b = 1)$, the expectation values of potential and kinetics based on quantities ‘a’ and ‘b’ are calculated as follows:
\bea
\ev{V}&=& \frac{e^{-2az}\pi(a^2B(3+2az+e^{2az}(-3 - 2 a z (-2 + a z)))}{2a^6},\\\nn&+&\frac{A (-15 + 3 e^{2az} (5 + a z (-4 + a z)) - 
    a z (18 + a z (9 + 2 a z))))}{2a^6}.
\eea
And 
\bea
\nn
\ev{T} &=& 4 \pi \int_0^\infty  \psi^*(r) * r^2 *T* \psi(r)  dr, \\\nn &=&\frac{e^{-2az} \pi(\sinh{az}+az(-\cosh{az}+az\sinh{az}))}{a^3} 
\eea
The expectation value of Hamiltonian is:
\bea
\ev{H} &=& \frac{e^{-2az} \pi(\sinh{az}+az(-\cosh{az}+az\sinh{az}))}{a^3} \\\nn &+&  \frac{e^{-2az}\pi(a^2B(3+2az+e^{2az}(-3 - 2 a z (-2 + a z)))}{2a^6},\\\nn&+&\frac{A (-15 + 3 e^{2az} (5 + a z (-4 + a z)) - 
    a z (18 + a z (9 + 2 a z))))}{2a^6}.
\eea
\bea
\nn
\frac{d}{da}\ev{H} &=& - \frac{(90A-60aAz+a^5 z^2+12a^2 (Az^2-B))\pi}{2a^7 } \\ \label{17} &-&  \frac{(2a^4 z(1+2zB)+3a^3 (1+4zB))\pi}{2a^3}.
\eea
To minimize the Hamiltonian, we used Mathematica 12.2 built-in function, called FindMinimium by assigning an initial guess value for the parameter $a$ from the plot of $H$ versus $a$ for each certain values of the input of the other model parameters. See Tables~(\ref{3a}-\ref{3d}) summarize the numerical results for the cases that are studied.

The wavefunction at origin (WFO) is found as:
\be
|\psi(0)|^2 = \frac{z^2}{ I}
\ee

For the second choice of the trial wavefunction as Harmonic oscillator wavefunction $(b = 2)$, the expected values of potential and kinetics based on quantities ‘a’ and ‘b’ are calculated as follows:
\bea
\ev{V} &=& \frac{\pi(-4 (A - a B) e^{-2 a z^2} + 4 (A + a A z^2 - a B (1 + 2 a z^2)) }{8 a^3}\\\nn&+&\frac{ \sqrt{a} (-3 A + 4 a B) \sqrt{2 \pi} z Erf[\sqrt{2} \sqrt{a} z])}{8 a^3}.
\eea
And 
\bea
\nn
\ev{T} &=& 4 \pi \int_0^\infty \psi^* r^2 T (r-z)e^{-a r^2} dr\\ &=&  \frac{\pi(4 \sqrt{a} (-8 + e^{-2 a z^2}) z + 
 \sqrt{2 \pi} (7 + 12 a z^2) Erf[\sqrt{2} \sqrt{a} z])}{32 a^{\frac{3}{2}}}
\eea
The expectation value of Hamiltonian is then
\bea
\ev{H} &=&  \frac{\pi(4 \sqrt{a} (-8 + e^{-2 a z^2}) z + 
 \sqrt{2 \pi} (7 + 12 a z^2) Erf[\sqrt{2} \sqrt{a} z])}{32 a^{\frac{3}{2}}}\\\nn&+&\frac{\pi(-4 (A - a B) e^{-2 a z^2} + 4 (A + a A z^2 - a B (1 + 2 a z^2)) }{8 a^3}\\\nn&+&\frac{  
 \sqrt{a} (-3 A + 4 a B) \sqrt{2 \pi} z Erf[\sqrt{2} \sqrt{a} z])}{8 a^3}.
\eea
We substitute the values of A and B as well as the values of the radius in differentiating $\ev{H}$, and set it equal to zero to find (a) values. Then we take one value for (a) for each radius and then substitute it in Eq. \eqref{6} and take into account that the absolute square of the complex part will be unity. 
\subsection{For the Global potential}
Similar procedure is applied here but with global potential. Using the Hydrogen wavefunction or exponential wavefunction $(b = 1)$ as a trial wavefunction,
the expected values of potential and kinetics based on quantities ‘a’ and ‘b’ are calculated as follows:
\bea
\ev{V} &=&\frac{ \pi}{512 a^6} [945 \sqrt{a} A \sqrt{2 \pi}-840 a^{\frac{3}{2}} A \sqrt{2 \pi} z - 512 a^4 B z^2 \\\nn&+&240 a^{\frac{5}{2}} A \sqrt{2 \pi} z^2-945 A \sqrt{2 a\pi} +512 a^3 e^{-2 a z} z (B + 2 B e^{2 a z} + (-1 + e^{2 a z}) C z - A z^{\frac{3}{2}})\\\nn&-&48 a^2 e^{-2 a z} (16 B (-1 + e^{2 a z}) + 32 (1 + e^{2 a z}) C z + 
   5 A z^{\frac{3}{2}} (7 +e^{-2 a z} \sqrt{2az \pi} ))
   \\\nn&+&12 a e^{-2 a z} (128 (-1 + e^{2 a z}) C + 
   35 A \sqrt{z} (-9 + 2 e^{2 a z} \sqrt{2\pi a z}))\\\nn&+&15 A\sqrt{2\pi a } (63 - 56 a z + 16 a^2 z^2) Erf[\sqrt{2\pi az }]]
\eea
And
\be
\ev{T} = \frac{e^{-az}\pi(\sinh{a z} + a z (-\cosh{a z} + a z \sinh{a z}))}{a^3}.
\ee

The expectation value of Hamiltonian is then:
\bea
\ev{H} &=&   \frac{e^{-az}\pi(\sinh{a z} + a z (-\cosh{a z} + a z \sinh{a z}))}{a^3}\\\nn&+&\frac{ \pi}{512 a^6} [945 \sqrt{a} A \sqrt{2 \pi}-840 a^{\frac{3}{2}} A \sqrt{2 \pi} z - 512 a^4 B z^2 \\\nn&+&240 a^{\frac{5}{2}} A \sqrt{2 \pi} z^2-945 A \sqrt{2 a\pi} +512 a^3 e^{-2 a z} z (B + 2 B e^{2 a z} + (-1 + e^{2 a z}) C z - A z^{\frac{3}{2}})\\\nn&-&48 a^2 e^{-2 a z} (16 B (-1 + e^{2 a z}) + 32 (1 + e^{2 a z}) C z + 5 A z^{\frac{3}{2}} (7 +e^{-2 a z} \sqrt{2az \pi} ))
   \\\nn&+&12 a e^{-2 a z} (128 (-1 + e^{2 a z}) C + 
   35 A \sqrt{z} (-9 + 2 e^{2 a z} \sqrt{2\pi a z}))\\\nn&+&15 A\sqrt{2\pi a } (63 - 56 a z + 16 a^2 z^2) Erf[\sqrt{2\pi az }]]
\eea
Differentiating $\ev{H}$ with respects to $(a)$.
We assume different values of the radius and substitute it in differentiating $\ev{H}$ and set it equal to zero to find $(a)$ values then we will take one value for $(a)$  for each radius and then substitute it in Eq. \eqref{6}. 

Repeating the same steps but with the Harmonic oscillator wavefunction $(b = 2)$ for the global potential, the expected values of potential and kinetics based on quantities ‘a’ and ‘b’ are calculated as follows:
\bea
\nn
\ev{V} &=& \frac{\pi}{64 a^{\frac{11}{4}}}[-64 a^{\frac{7}{4}} B z^2 - 32 a^{\frac{7}{4}} (B + 2 C z)-10 2^{\frac{3}{4}}\sqrt{a}A z \Gamma{(\frac{1}{4}})\\\nn&+& 4 a^{\frac{1}{4}}\sqrt{2 \pi} (8 a B z + C (3 + 4 a z^2)) Erf[\sqrt{2az}]
\\\nn&+& 21 A 2^{\frac{1}{4}} \Gamma{(\frac{3}{4})} + 24 A2^{\frac{1}{4}} a  z^2 \Gamma{(\frac{3}{4})}]\\\nn&+& a^{\frac{3}{4}}8 e^{-2 a z^2} (4 B + 2 K z + 3 A z^{\frac{3}{2}}) +  10\sqrt{a}
   2^{\frac{3}{4}} A z \Gamma{(\frac{1}{4}, 2 a z^2)} 
   \\\nn&-& (
   3 2^{\frac{1}{4}} A (7 + 8 a z^2) \Gamma{(\frac{3}{4}, 2 a z^2)}).            
\eea
And
\be
\ev{T} =  \frac{\pi (4 \sqrt{a} (-8 + e^{-2 a z^2}) z + 
 \sqrt{2 \pi} (7 + 12 a z^2) Erf[\sqrt{2a z}])}{32 a^{\frac{3}{2}}}.
\ee
The expectation value of Hamiltonian:
\bea
\ev{H} &=&  \frac{\pi (4 \sqrt{a} (-8 + e^{-2 a z^2}) z + 
 \sqrt{2 \pi} (7 + 12 a z^2) Erf[\sqrt{2a z}])}{32 a^{\frac{3}{2}}}\\\nn&+&\frac{\pi}{64 a^{\frac{11}{4}}}[-64 a^{\frac{7}{4}} B z^2 - 32 a^{\frac{7}{4}} (B + 2 C z)-10 2^{\frac{3}{4}}\sqrt{a}A z \Gamma{(\frac{1}{4}})\\\nn&+& 4 a^{\frac{1}{4}}\sqrt{2 \pi} (8 a B z + C (3 + 4 a z^2)) Erf[\sqrt{2az}]
\\\nn&+& 21 A 2^{\frac{1}{4}} \Gamma{(\frac{3}{4})} + 24 A2^{\frac{1}{4}} a  z^2 \Gamma{(\frac{3}{4})}]\\\nn&+& a^{\frac{3}{4}}8 e^{-2 a z^2} (4 B + 2 K z + 3 A z^{\frac{3}{2}}) +  10\sqrt{a}
   2^{\frac{3}{4}} A z \Gamma{(\frac{1}{4}, 2 a z^2)} 
   \\\nn&-& (
   3 2^{\frac{1}{4}} A (7 + 8 a z^2) \Gamma{(\frac{3}{4}, 2 a z^2)}).   
\eea
We assume different values of the radius and substitute it in differentiating $\ev{H}$ and set it equal to zero to find $(a)$ values then we will take one value for $(a)$ for each radius and then substitute it in Eq (6). 
\section{Results and Discussion}
We present our numerical calculations for the two potential models using two different trial wavefunctions at different cut-off radii. The behaviour of the probability densities of these trial wavefunctions are shown in Figure~(\ref{3ab}) for distinct parameters.
\begin{figure}[!htbp]
  \centering
  \subfigure [$A=0.5, B=2, b=1$] {\includegraphics[scale=0.62]{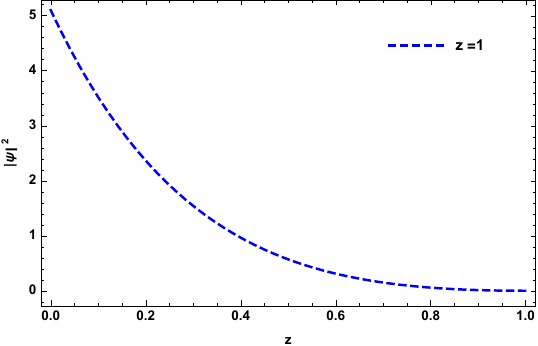}}
  \subfigure [$A=0.5, B=2,b=1$]{\includegraphics[scale=0.62]{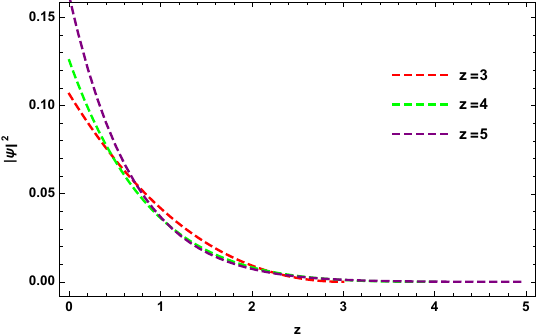}}
   \subfigure [$A=0.5, B=2,C=0.8, b=1$] {\includegraphics[scale=0.62]{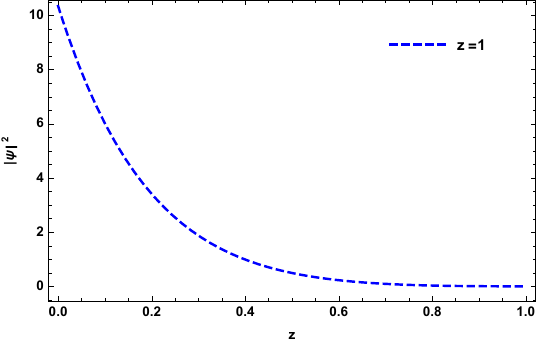}}
  \subfigure [$A=0.5, B=2,C=0.8,b=1$]{\includegraphics[scale=0.62]{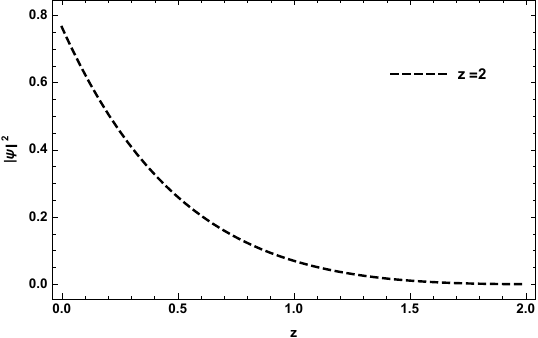}}
   \subfigure [$A=0.5, B=2,C=0.8, b=1$] {\includegraphics[scale=0.62]{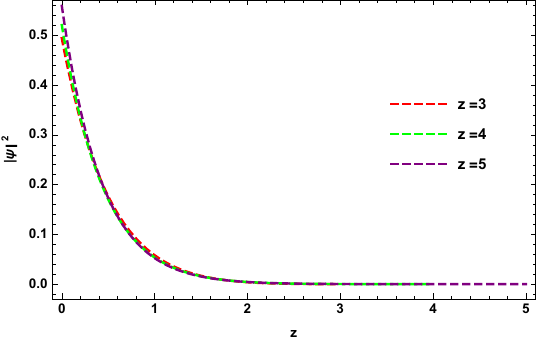}}
  \hfill
  \caption {$|\psi(r)|^2$ for Cornell potential (upper panel) and global potential (lower panel) at different cut-off radii $z$.}
	\label{3ab}
\end{figure}\newpage
\begin{figure}[!htbp]
  \centering
  \subfigure [$A=0.5, B=2, b=2$] {\includegraphics[scale=0.62]{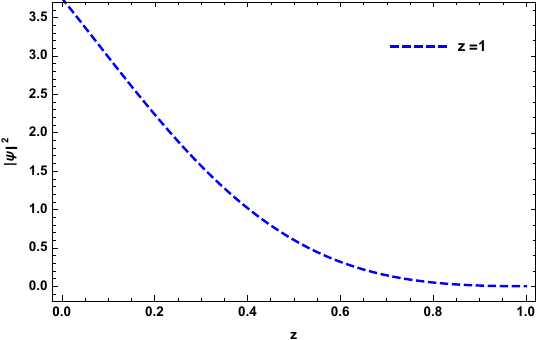}}
  \subfigure [$A=0.5, B=2,b=2$]{\includegraphics[scale=0.62]{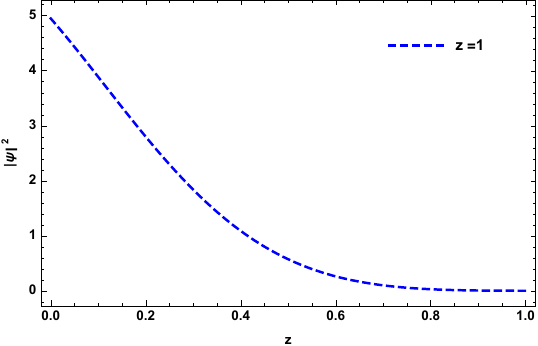}}
   \subfigure [$A=0.5, B=2,C=0.8, b=2$] {\includegraphics[scale=0.62]{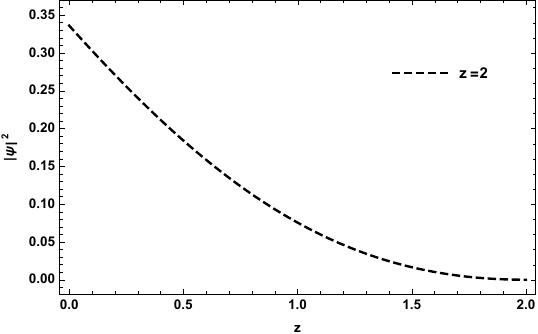}}
  \subfigure [$A=0.5, B=2,C=0.8,b=2$]{\includegraphics[scale=0.62]{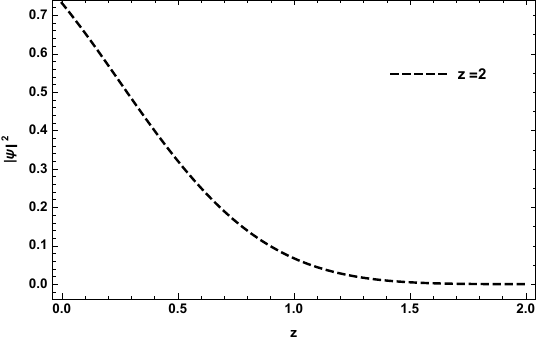}}
   \subfigure [$A=0.5, B=2,C=0.8, b=2$] {\includegraphics[scale=0.62]{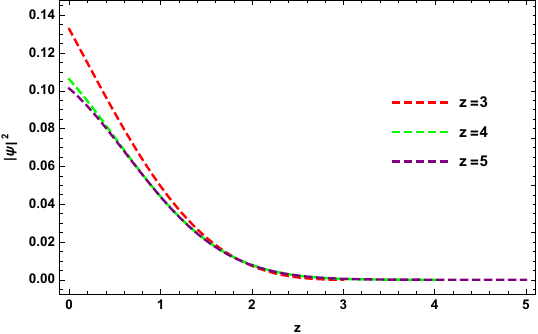}}
   \subfigure [$A=0.5, B=2,C=0.8, b=2$] {\includegraphics[scale=0.62]{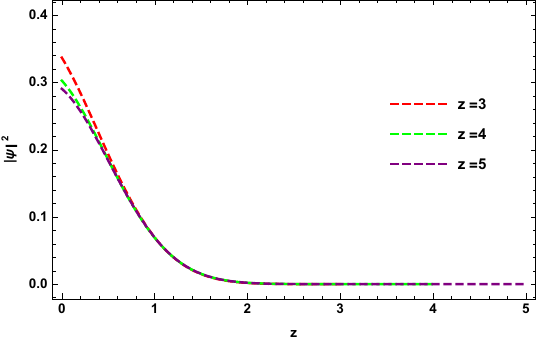}}
  \hfill
  \caption {$|\psi(r)|^2$ for Cornell potential (left panel) and global potential (right panel) at different cut-off radii $z$.}
	\label{3ab}
\end{figure}\newpage
Using these probability densities, other important physical quantities such as $|\psi(0)|^2$, $\ev{r}$, $E$ were calculated.

The wave function at the origin (WFO) is an important quantity while investigating various physical problems for some bound states. For example, it is needed to calculate the production and decay amplitudes of heavy quarkonia.
We present our results for the WFO as well as for the mean radius for confined systems under two potential models and various cut-off radii. We have used different trial wavefunctions for our calculations, namely, the 1S hydrogen radial wavefunction and the 1S SHO radial wavefunction

Tables~(\ref{3a},\ref{3b}) summarize the results for the Cornell potential, while Tables~(\ref{3c},\ref{3d}) for the global potential, for various possible cut-off radii. 
\begin{table}[htb!]
    \centering
    \caption{The variational results using the hydrogen wavefunction for the Cornell potential with $A=0.5$ and $B=2$.}
    \label{3a}
    \begin{tabular}{c|c|c|c|c}
         \textbf{z}& \textbf{a value} & $\boldsymbol{|\psi(0)|^2}$ & $\boldsymbol{\ev{r}}$(GeV) & \textbf{E (GeV)}  \\ \hline
\textbf{1} &	0.804769	&5.09835	&0.443098	&-0.050878 \\ \hline
\textbf{3} &	0.0627947	&0.106478	&1.45965	&  -37.4639\\ \hline
\textbf{4}&	0.335031&	0.125545	&1.62754&-57.2779  \\ \hline
\textbf{5} &	0.519695	&0.162228	&1.65681 &-76.5087
    \end{tabular}
\end{table}
\begin{table}[hbt!]
    \centering
     \caption{The variational results using the harmonic oscillator wavefunction for the Cornell potential with $A=0.5$ and $B=2$.}
    \label{3b}
    \begin{tabular}{c|c|c|c|c}
         \textbf{z}& \textbf{a value} & $\boldsymbol{|\psi(0)|^2}$ & $\boldsymbol{\ev{r}}$(GeV) & \textbf{E (Gev)}  \\ \hline
\textbf{1} &	0.885092&	3.75142&	0.4424&-0.0547323	 \\ \hline
\textbf{2} &	0.0533772&	0.336013	&0.970131	& -10.1952 \\ \hline
\textbf{3} &	0.0870008&	0.132553&	1.34537	& -43.0644 \\ \hline
\textbf{4}&0.15255&0.106019&	1.46991&	-87.0158 \\ \hline
\textbf{5} &	0.194375&	0.101128&	1.48399	&-141.646
    \end{tabular}
\end{table}
\begin{table}[hbt!]
    \centering
    \caption{The variational results using the exponential wavefunction for the global potential with $A=0.5$, $B=2$ and $C=0.8$}
    \label{3c}
    \begin{tabular}{c|c|c|c|c}
         \textbf{z} & \textbf{a value} & $\boldsymbol{|\psi(0)|^2}$ & $\boldsymbol{\ev{r}}$ (GeV)& \textbf{E (GeV)}  \\ \hline
         \textbf{1} &	1.65865& 	10.344	&0.386356&0.07915	 \\ \hline
\textbf{2} &	0.507205& 0.765083	&0.85739&-2.16297\\ \hline
\textbf{3} &	0.665471& 	0.493858	&1.09675	& -6.68171\\ \hline
\textbf{4} &	0.848832&	0.520176	&1.16739&-11.3696	 \\ \hline
\textbf{5}&	0.960957	&0.55883	&1.18452&-16.993	 
    \end{tabular}
\end{table}
\begin{table}[hbt!]
    \centering
    \caption{The variational results using the harmonic oscillator wavefunction f\ $^3$Ministry of Education and Higher Education, Qataror the global potential  with $A=0.5$, $B=2$ and $C=0.8$}
    \label{3d}
    \begin{tabular}{c|c|c|c|c}
         \textbf{z}& \textbf{a value} & $\boldsymbol{|\psi(0)|^2}$ & $\boldsymbol{\ev{r}}$ (GeV)& \textbf{E (GeV)}  \\ \hline
         \textbf{1} &	1.54294& 4.94705&0.40693	& 0.157745\\ \hline
\textbf{2} &	0.5&	0.731946&	0.771337&-2.61122	\\ \hline
\textbf{3} &	0.383842&	 0.336937&1.13758&-9.34193	 \\ \hline
\textbf{4} &	0.452373	&0.302755&	1.02107&-17.6167	 \\ \hline
\textbf{5}&	0.493727	&0.290524&	1.022&	-28.0771 
    \end{tabular}
\end{table}\newpage
\begin{figure}[!htb]
  \centering
  \subfigure [$A=0.5, B=2,b=1$]{\includegraphics[scale=0.62]{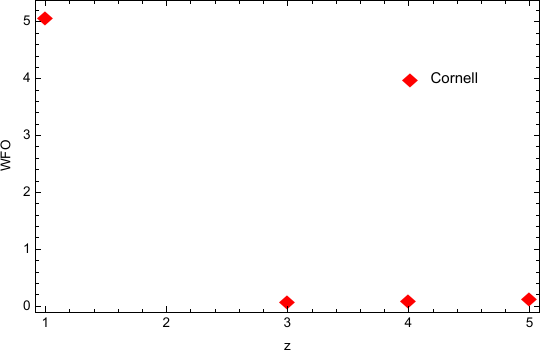}}
	\subfigure [$A=0.5, B=2,C=0.8, b=1$] {\includegraphics[scale=0.62]{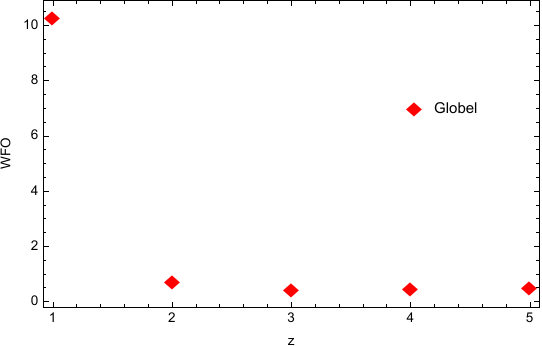}}
  \subfigure [$A=0.5, B=2,C=0.8,b=2$]{\includegraphics[scale=0.62]{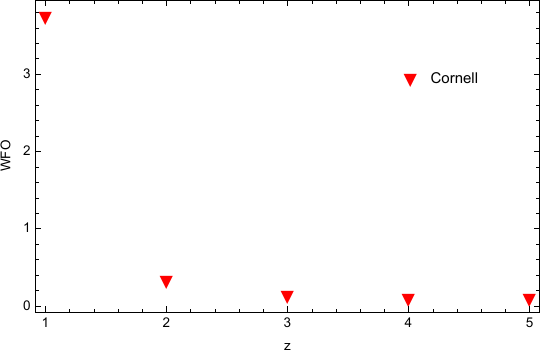}}
  \subfigure [$A=0.5, B=2,C=0.8,b=2$]{\includegraphics[scale=0.62]{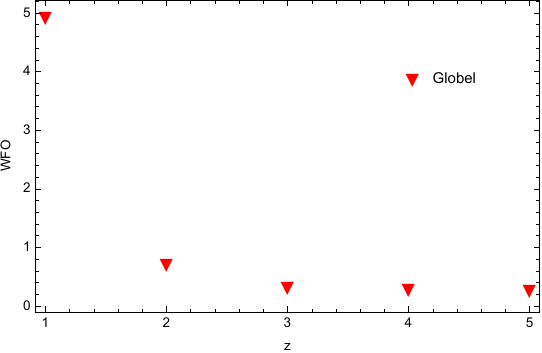}}
  \hfill
  \caption {WFO for Cornell potential (left panel) and global potential (right panel) at different cut-off radii $z$.}
	\label{3abc}
\end{figure}
\section{Conclusions}
In this work, we have calculated some properties of hardly confined two-particle systems with two potential models, namely, the Cornell potential and the global potential. The variational method is used with trial wavefunctions of the form $1S$ state hydrogen-like wavefunction, or a $1S$ harmonic oscillator-like wavefunction, times a cut-off function of the form $(r-z)$. With these two trial wavefunctions, the quantities  $|\psi|^2$, WFO, the energies $E$ as well as the mean radius $(r)$ were obtained for the two potential models at different cut-off radii $z$. As shown in the produced figures in the previous section, there is generally a similar behavior in the calculated quantities for the two potential models for the input potential parameters.  The general behaviors are similar to these reported in \cite{Boroun_2009}, for strongly confined systems, such as heavy quarkonia.  We may infer that each of the discussed model, with suitable choices of potential parameters, could be applicable for investigating hardly confined subjected to Dirichlet boundary conditions. Higher levels states could be also studied by extended the trial wavefunction to include higher degree polynomials as cut-off functions. This will be of our future work.
\bibliography{ref} 
\bibliographystyle{unsrtnat}
 \pdfoutput=1
\end{document}